\documentclass[aps,jcp,twocolumn,superscriptaddress,floatfix]{revtex4}

\usepackage{color}
\usepackage{dcolumn} 
\usepackage{bm} 
\usepackage{graphicx}
\usepackage{multirow} 
\usepackage{pifont} 
\usepackage{epsfig}
\usepackage{verbatim}
\usepackage{subfigure}
\usepackage{float}
\usepackage{pdflscape}
\usepackage{amsmath,amsfonts,amssymb,bm,graphicx,color,braket}
\usepackage[version=4]{mhchem}
\usepackage{array}
\usepackage[columnwise]{lineno}
\usepackage{mathrsfs}
\usepackage[nolist]{acronym}
\usepackage[mathcal]{euscript}
\usepackage[titletoc,toc,title]{appendix}
\usepackage{textcomp}
\usepackage{threeparttable}
\usepackage{enumitem}
\usepackage[utf8]{inputenc}

\newcommand{\etal}{\emph{et~al.}}

\newcommand{\al}{\alpha}
\newcommand{\be}{\beta}
\newcommand{\de}{\delta}
\newcommand{\De}{\Delta}

\newcommand{\ze}{\zeta}
\newcommand{\om}{\omega}

\newcommand{\sig}{\sigma}

\newcommand{\br}{\mathbf{r}}

\newcommand{\EXC}{E_\text{XC}}

\newcommand{\zetr}{\ze_\text{t}}
\newcommand{\zeftr}{\ze_\text{ft}}

\newcommand{\EZPV}{\De E_\text{S-T}^\text{ZPVE}}
\newcommand{\EST}{\De E_\text{S-T}}

\newcommand{\EMCPDFT}{E_\text{MC-PDFT}}
\newcommand{\EOTPD}{E_\text{OTPD}}

\newcommand{\mc}{\multicolumn}
\newcommand{\mr}{\multirow}

\newcommand{\alert}[1]{\textcolor{red}{#1}}

\setenumerate[1]{label=\thesection.\arabic*.}
\begin{document}

\title{Combining pair-density functional theory and variational two-electron reduced-density matrix methods}

\author{Mohammad Mostafanejad}
\affiliation{
             Department of Chemistry and Biochemistry,
             Florida State University,
             Tallahassee, FL 32306-4390}
\author{A. Eugene DePrince III}
\affiliation{
             Department of Chemistry and Biochemistry,
             Florida State University,
             Tallahassee, FL 32306-4390}
\email{deprince@chem.fsu.edu}



\begin{abstract}

Complete active space self-consistent field (CASSCF) computations can be
realized at polynomial cost via the variational optimization of the
active-space two-electron reduced-density matrix (2-RDM).  Like
conventional approaches to CASSCF, variational 2-RDM (v2RDM)-driven CASSCF
captures nondynamical electron correlation in the active space, but it
lacks a description of the remaining dynamical correlation effects.  Such
effects can be modeled through a combination of v2RDM-CASSCF and on-top
\ac{PDFT}.  The resulting v2RDM-CASSCF-PDFT approach provides a
computationally inexpensive framework for describing both static and
dynamical correlation effects in multiconfigurational and strongly
correlated systems.  On-top pair-density functionals can be derived from
familiar Kohn-Sham \ac{XC} density functionals through the translation of
the v2RDM-CASSCF reference densities 
[Li Manni \etal, J. Chem. Theory Comput. {\bf 10}, 3669-3680 (2014)].
Translated and fully-translated on-top PDFT versions
of several common \ac{XC} functionals are applied to the \aclp{PEC} of
N$_2$, H$_2$O, and CN$^{-}$, as well as to the singlet/triplet energy
splittings in the linear polyacene series.  Using v2RDM-CASSCF-PDFT and
the translated PBE functional, the singlet/triplet energy splitting of an
infinitely-long acene molecule is estimated to be 4.87 kcal mol$^{-1}$.





\end{abstract}

\maketitle

\section{Introduction}
\label{SEC:INTRODUCTION}

The accurate and computationally affordable description of the electronic
structure of many-body systems remains a major challenge within the
quantum chemistry and molecular physics
communities.\cite{Kutzelnigg2003Ch,Kutzelnigg1973,Hattig2012}
Specifically, the realization of general approaches that account for both
dynamical and nondynamical correlation effects in multiconfigurational
\cite{Roos2016} or strongly-correlated systems is particularly difficult.
The main issue is that many approaches designed to deal with \ac{MR}
problems are not particularly efficient for capturing dynamical correlation
effects.  A similar statement can be made regarding the ability of methods
designed to model dynamical correlation to capture \ac{MR} effects.


One can broadly classify approaches to the electron correlation problem as
either falling within \ac{WFT}, in which the many-electron wave function
is obviously the central quantity,
\cite{Helgaker_book,Szabo1996,Yarkony1995} or density-based theories,
which include both \ac{DFT}
\cite{Parr1989,Fiolhais2003,Garza2016,Cohen:2012:289,Becke2014} and
\ac{RDM}-based approaches.
\cite{Cioslowski2000,Davidson1976,Coleman2000,Mazziotti2007,Gidopoulos2003}
In principle, \ac{WFT} is preferable, as it allows for systematic
improvement in the calculated energies and properties of the
system.\cite{Helgaker2012} In practice, however, the computational
complexity of post-\acl{HF} wave-function-based methods, specifically
\ac{MR} approaches, limits their application to small
systems.\cite{Szalay2012} The wide-ranging success of \ac{DFT}, on the other
hand, stems from the its ability to provide a reasonable description of
electron correlation at significantly lower costs.  Nonetheless, \ac{DFT}
often fails for \ac{MR} systems, and it does not offer a systematic
approach for improving its accuracy.
\cite{Parr1989,Fiolhais2003,Garza2016,Cohen:2012:289,Becke2014}


Within WFT, one of the most familiar approaches to the \ac{MR} problem is
the \ac{CASSCF} method \cite{Roos:1980:157, Siegbahn:1980:323,
Siegbahn:1981:2384, roos:casscf}.  In \ac{CASSCF}, the molecular orbitals
are partitioned into inactive (doubly occupied), active (partially
occupied), and external (empty) orbitals, and the active space is chosen
with some knowledge as to which orbitals are important for the chemical
problem at hand.  In the canonical form of CASSCF, the active-space
electronic structure is described by a full \ac{CI} wave function, and it
is assumed that all nondynamical correlation effects are captured by this
procedure.  Dynamical correlation effects can then be incorporated through
a variety of approaches, including perturbation theory (using, for
example, \ac{CASPT2} \cite{Andersson1990,Andersson1992}). \ac{CASPT2}
requires knowledge of the \ac{4-RDM}, or some approximation to it, which
can become problematic as the size of the active space increases.
Accordingly, several approaches have been proposed that eliminate the
manipulation of the \ac{4-RDM}, including the \ac{ACSE}
\cite{Mukherjee:2001:2047,Kutzelnigg:2004:7350,Mazziotti:2006:143002} and
the \ac{DSRG} \cite{Evangelista2014}.  Both of these approaches require
knowledge of the \ac{3-RDM}.


A long sought-after alternative to the methods described above involves
the combination of the \ac{MR} approach to the static correlation problem
with a \ac{DFT}-based description of dynamical correlation.  Since the
advent of the \ac{MR}+\ac{DFT} framework,
\cite{Lie:1275:1974,Lie:1288:1974} a substantial amount of effort has been
devoted to increasing the accuracy and efficiency of this approach.
\cite{Garza2015,Perez2007,Colle1979,Moscardo1991,Garza2013,Paziani2006,Gusarov2004,Gusarov2004a,Takeda2002,Stoll2003,Grafenstein2000,Grafenstein2005,MIEHLICH1997,Grimme1999,Wu1999,Pollet2002,Leininger1997,Malcolm1996,Kraka1992}
There are several issues that one must carefully consider when developing
a \ac{MR}+\ac{DFT} scheme, including (i) the symmetry dilemma that plagues
\ac{KS}-\ac{DFT} in general, (ii) the double counting of electron
correlation within the active space, and (iii) the steep computational
scaling of many commonly used \ac{MR} methods.  The framework of the
\ac{MC-PDFT} \cite{Gagliardi2017,LiManni2014} addresses the first two
issues while leaving open the question of the cost of the evaluation of
the underlying \ac{MR} wave function.  The success of the original
formulation of \ac{MC-PDFT}\cite{LiManni2014} notwithstanding, Garza \etal
\cite{Garza2015} rightly note that, if the \ac{MR} component is determined
using an approach such as \ac{CASSCF}, its utility is potentially limited
by the exponential complexity of the \ac{CI}-based active-space wave
function.  Hence, those authors propose that the static correlation within
\ac{MC-PDFT} be described by the \ac{pCCD} method,\cite{Limacher:2013:1394,Stein:2014:214113} the scaling of which
increases as only $\mathcal{O}(k^4)$ or $\mathcal{O}(k^5)$,
where $k$ is the size of the one-electron basis set; the specific
scaling is determined by the threatment of the orbital transformation step.
More recently, the MC-PDFT
scheme has also been employed in conjunction with other active-space-based
methods that scale more favorably than CASSCF, including the
\ac{DMRG}\cite{Sharma:2018:arXiv} and the
\ac{GASSCF}\cite{Ghosh:2017:2741}.

Here, we offer an alternative strategy to overcome the problematic scaling
of \ac{CI}-based \ac{CASSCF} within the \ac{MC-PDFT} framework.  We elect
to maintain the CASSCF-based description of the static correlation problem
utilized in the original formulation of the approach,\cite{LiManni2014}
but we represent the electronic structure of the active space with the
\ac{2-RDM}, as opposed to the \ac{CI} wave function.  The computational
complexity of 
\ac{v2RDM-CASSCF} \cite{Gidofalvi:2008:134108,FossoTande:2016:2260} increases
only polynomially with the size of the active space, thereby facilitating
\acs{v2RDM}-based \ac{MC-PDFT} computations (denoted \acs{v2RDM-CASSCF-PDFT})
on active spaces as large as 50 electrons in 50 orbitals.

This paper is organized as follows.  Section \ref{SEC:THEORY} provides an
overview of \acs{v2RDM-CASSCF-PDFT}, including brief discussions of the
theory underlying the \ac{v2RDM-CASSCF} and \ac{MC-PDFT} schemes.  The
computational details of the work are then given in Sec.
\ref{SEC:COMPDETAILS}.  In Sec.  \ref{SEC:RESULTS}, we provide evidence
of the correctness of our \acs{v2RDM-CASSCF-PDFT} implementation by
comparing singlet/triplet energy splittings for a set of main-group divalent radicals
to those obtained from conventional, CI-CASSCF-driven MC-PDFT.\cite{Bao2016}  We then
apply
\acs{v2RDM-CASSCF-PDFT} to the \acp{PEC} of N$_2$, H$_2$O, and CN$^-$, as
well as to the singlet/triplet energy gaps of the linear polyacene series.
Some concluding remarks can be found in Sec. \ref{SEC:CONCLUSIONS}.

\section{Theory}
\label{SEC:THEORY}

Throughout this work, we adopt the conventional notation employed within
\ac{MR} methods for labeling molecular orbitals (MOs, $\lbrace \psi
\rbrace$): the indices $i$, $j$, $k$, and $l$ represent inactive orbitals;
$t$, $u$, $v$, and $w$ indicate active orbitals; $a$, $b$, $c$, and $d$
denote external orbitals; and $p$, $q$, $r$, and $s$ represent general
orbitals.


Let $\Psi$ be an $N$-electron wave function in Fock space.
One- and two- particle excitation operators can be expressed as
\cite{Helgaker_book}
\begin{subequations}
	\begin{gather}
	\hat{E}^p_q = \hat{a}^\dag_{p_\sig} \hat{a}_{q_\sig}   \label{EQ:Epq}  	\\	
	\hat{e}^{p r}_{q s} = \hat{E}^p_q \hat{E}^r_s - \de^q_r \hat{E}^p_s = \hat{a}^\dag_{p_\sig} \hat{a}^\dag_{r_\tau} \hat{a}_{s_\tau} \hat{a}_{q_\sig}		\label{EQ:epqrs}
	\end{gather}
\end{subequations}
where $\hat{a}^\dag$ and $\hat{a}$ represent second-quantized creation and
annihilation operators, respectively, and the Greek labels run over $\al$ and
$\be$ spins. Einstein's summation convention is implied throughout. The
non-relativistic \ac{BO} electronic Hamiltonian is
\begin{equation}
\hat{\mathscr{H}} = h^p_q \hat{E}^p_q + \frac{1}{2} \nu^{pq}_{rs} \hat{e}^{pq}_{rs}
\end{equation}
where $h^p_q = \braket{\psi_p|\hat{h}|\psi_q} $ is the sum of the electron
kinetic energy and electron-nucleus potential energy integrals, and
$\nu^{pq}_{rs} = \braket{\psi_p \psi_q|\psi_r\psi_s}$ represents the
two-electron repulsion integral tensor. Because the electronic Hamiltonian
includes up to only pair-wise interactions, the ground-state energy of a
many-electron system can be expressed as an exact linear functional of the
\ac{2-RDM} and the \ac{1-RDM}
\cite{Husimi:1940:264,Mayer:1955:1579,Lowdin:1955:1474}
\begin{equation}\label{EQ:Eel}
E = {}^1D^p_q h^p_q + \frac{1}{2} {}^2D^{pq}_{rs} \nu^{pq}_{rs}.
\end{equation}
Here, the \ac{1-RDM} and the \ac{2-RDM} are represented in their spin-free forms, 
defined as
\begin{equation}
{}^1D^p_q = {}^1D^{p_\sigma}_{q_\sigma} = \braket{\Psi|\hat{E}^p_q|\Psi}	\label{EQ:1RDM}		\\
\end{equation}
and
\begin{equation}
{}^2D^{pq}_{rs} = {}^2D^{p_\sigma q_\tau}_{r_\sigma s_\tau} = \braket{\Psi|\hat{e}^{pq}_{rs}|\Psi} \label{EQ:2RDM}.
\end{equation}
Summation over the spin labels is implied.

\subsection{v2RDM-driven CASSCF}\label{SUBSEC:N_REP}

The \ac{CASSCF} non-relativistic \ac{BO} electronic Hamiltonian is
\begin{equation}
\hat{\mathscr{H}}_{\text{CASSCF}} = (h^t_u + 2\nu^{ti}_{ui} - \nu^{tu}_{ii}) \hat{E}^t_u + \frac{1}{2} \nu^{tu}_{vw} \hat{e}^{tu}_{vw},
\end{equation}
and the \ac{CASSCF} active-space electronic energy is expressible in terms of the active-space 1-
and \acp{2-RDM}
\begin{equation}\label{EQ:ECAS}
E_{\text{CASSCF}} = (h^t_u + 2\nu^{ti}_{ui} - \nu^{tu}_{ii}) {}^1D^t_u + \frac{1}{2} \nu^{tv}_{uw} {}^2D^{tv}_{uw}.
\end{equation}
The central idea of \ac{v2RDM-CASSCF} is that the spin blocks of the
active-space RDMs can be determined directly by minimizing the energy with
respect to variations in their elements (and to variations in the orbital
parameters).\cite{Gidofalvi:2008:134108,FossoTande:2016:2260}
Because not every \ac{2-RDM} can be derived from an $N$-electron wave
function, this procedure can lead to unphysically low
energies;\cite{Cioslowski2000} a physically meaningful \ac{2-RDM} should
fulfill certain $N$-representability conditions.  These conditions are
most easily expressed in terms of the individual spin blocks that contribute
to the spin-free \ac{2-RDM}.
Specifically, each
spin block of the \ac{2-RDM} should 
(i) be Hermitian,
(ii) be antisymmetric with respect to the permutation of particle labels,
(iii) conserve the number of pairs of particles (have a fixed trace),
and (iv) contract to the appropriate spin block(s) of the
\ac{1-RDM}.
Constraints on the expectation value of $\hat{S}^2$ can also be
applied.\cite{Perez-Romero1997}

In addition to these trivial constraints on the \ac{2-RDM}, all
spin blocks of all \acp{RDM} should be positive semidefinite.  Such
positivity conditions applied to the spin blocks of the \ac{2-RDM}, the
two-hole \acs{RDM}, and the particle-hole \acs{RDM} constitute the
two-body (PQG) constraints of Garrod and Percus.\cite{Garrod:1964:1756}
Additional positivity conditions can be applied to higher-order \acp{RDM}.
In this work, we consider the PQG constraints as well as the T2 partial
three-particle condition.\cite{Zhao2004,Erdahl:1978:697}  The spin blocks
of each of these RDMs are interrelated through linear mappings implied by the
anticommutation properties of the creation and annihilation operators that
define them.  The \ac{v2RDM-CASSCF} procedure thus involves a large-scale
semidefinite optimization that we carry out using a boundary-point
algorithm \cite{Povh:2006:277,Malick:2009:336,Mazziotti:2011:083001}, the
specific details of which can be found in Ref.
\citenum{FossoTande:2016:2260}.

\subsection{Multi-configuration Pair-Density Functional Theory}\label{SUBSEC:MCPDFT}


One of the main pitfalls of the \ac{MR}+\ac{DFT} scheme is the double
counting of electron correlation within the active space. The most
expedient solution is to employ a small active space or modified
functionals;\cite{Lie:1275:1974,Lie:1288:1974,Perez2007} such strategies
may not always lead to satisfactory results, though.  A seemingly robust
solution\cite{LiManni2014} partitions the interelectronic Coulomb
contribution to the energy into a classical Coulomb component (obtained
from an \ac{MR} method) and all other exchange and pure two-electron
contributions (described by DFT).  Because the two-electron correlations
are modeled entirely within the framework of DFT, double counting of such
contributions to the energy is automatically avoided.  However, this
strategy offers no such guarantee regarding the double counting of {\em
kinetic} correlation.

A second complication in \ac{MR}+\ac{DFT} is related to the ``symmetry
dilemma'' \cite{Perdew1995} of standard \ac{KS}-\ac{DFT}.  One manifestion
of this issue within \ac{MR}+\ac{DFT} is the incompatibility of standard
\ac{XC} functionals with \ac{MR} spin densities for
low-spin (i.e. $|M_S| < S$) states.  Fortunately, this difficulty is
easily overcome by replacing the usual independent variables that enter
\ac{KS}-\ac{DFT} \ac{XC} functionals, the total density, $\rho(\br) =
\rho_\alpha(\br) + \rho_\beta(\br)$, and the spin magnetization, $m(\br) =
\rho_\alpha(\br) - \rho_\beta(\br)$, with the total density and the
\ac{OTPD}, $\Pi(\br)$.\cite{Becke1995,Moscardo1991,Perdew1995}

Both double counting and the symmetry dilemma are addressed
through the framework of MC-PDFT,\cite{LiManni2014} in which the 
active-space electronic energy is defined as
\begin{eqnarray}
\label{EQ:EMCPDFT}
	\EMCPDFT = (h^t_u + 2\nu^{ti}_{ui}) {}^1D^t_u + \frac{1}{2} \nu^{tv}_{uw} {}^1D^{t}_{u} {}^1D^{v}_{w} \nonumber \\
             + \EOTPD\left[\rho(\br), \Pi(\br), |\nabla\rho(\br)|, |\nabla\Pi(\br)| \right].
\end{eqnarray}
Here, the two-electron term from Eq.~\ref{EQ:ECAS} has been replaced by a
classical Coulombic term, and the remaining exchange and correlation
effects are folded into a functional of the \ac{OTPD}.  The total
electronic density and its gradient are defined by the \ac{1-RDM} as
\begin{equation}
 	\label{EQ:RHO}	
\rho(\br) = {}^1D^p_q\ \psi^*_p(\br) \psi_q(\br),
\end{equation}
and
\begin{equation}
\label{EQ:DRHO}
\nabla\rho(\br) = {}^1D^p_q \left[ \nabla\psi^*_p(\br) \psi_q(\br) + \psi^*_p(\br) \nabla\psi_q(\br) \right],
\end{equation}
respectively. 
The \ac{OTPD} and its gradient can similarly be defined in
terms of the \ac{2-RDM} as
\begin{equation}
\label{EQ:PI}
\Pi(\br)  = {}^2D^{pq}_{rs}\ \psi^*_p(\br) \psi^*_q(\br) \psi_r(\br) \psi_s(\br),
\end{equation}
and
\begin{eqnarray}
\label{EQ:DPI}	
\nabla\Pi(\br) = {}^2D^{pq}_{rs} &[& \nabla\psi^*_p(\br) \psi^*_q(\br) \psi_r(\br) \psi_s(\br) \nonumber \\
                                 &+& \psi^*_p(\br) \nabla\psi^*_q(\br) \psi_r(\br) \psi_s(\br) \nonumber \\
                                 &+& \psi^*_p(\br) \psi^*_q(\br) \nabla\psi_r(\br) \psi_s(\br) \nonumber \\
                                 &+& \psi^*_p(\br) \psi^*_q(\br) \psi_r(\br) \nabla\psi_s(\br) ~],
\end{eqnarray}
respectively. Here, the 1- and 2-RDMs are obtained from an MR computation.

Armed with a potentially robust framework for \ac{MR}+\ac{DFT}, we must
identify a suitable \ac{OTPD} functional for use within \ac{MC-PDFT}.  The
simplest class of functionals can be derived from existing approximate
\ac{XC} functionals employed within \ac{KS}-\ac{DFT} by first recognizing
that, for a density derived from a single Slater determinant, the spin
magnetization can be expressed exactly in terms of the \ac{OTPD} and the
total density.\cite{Moscardo1991,Becke1995} More specifically, the spin
polarization factor, $\ze(\br) = m(\br)/\rho(\br)$, can be expressed as
\begin{equation}
\label{EQ:ZETR}  
    \ze(\br) = \sqrt{1-R(\br)},
\end{equation}
where
\begin{equation}
\label{EQ:OTR}
    R(\br) = \frac{4~ \Pi(\br)}{\rho^2(\br)}.
\end{equation}
The basic assumption underlying the ``translated'' (t) \ac{OTPD}
functionals proposed in Ref. \citenum{LiManni2014} is that the spin
polarization factor can be similarly defined for a density and \ac{OTPD}
obtained from a \ac{MR} method, as
\begin{align}\label{EQ:TR}
        \zetr(\br) &= \begin{cases}
            \sqrt{1-R(\br)} &  \quad R(\br) \leq 1     \\[8pt]
            0               &  \quad R(\br) > 1       
    \end{cases} 
\end{align}
where the second case accounts for the fact that the argument of the
square root can become negative for $\rho(\br)$ and $\Pi(\br)$ that are
not derived from a single-configuration wave function.  The
translated \ac{OTPD} functional is then defined as
\begin{eqnarray}
\label{EQ:TRE}
\EOTPD\left[\rho(\br), \Pi(\br),| \nabla\rho(\br)| \right] \equiv \nonumber \\
\EXC[\tilde{\rho}_\al(\br), \tilde{\rho}_\be(\br), |\nabla\tilde{\rho}_\al(\br)|, |\nabla\tilde{\rho}_\be(\br)|],
\end{eqnarray}
where the tilde refers to translated densities and their gradients, given
by \cite{LiManni2014, Gagliardi2017}
\begin{equation}
\label{EQ:TR}
\tilde{\rho}_\sig(\br) = \frac{\rho(\br)}{2} \left(1 + c_\sig \zetr(\br)\right),
\end{equation}
and
\begin{equation}
\nabla\tilde{\rho}_\sig(\br) = \frac{\nabla\rho(\br)}{2} \left(1 + c_\sig \zetr(\br)\right),
\end{equation}
respectively.  Here, $c_\sig$ = 1~(-1) when $\sig = \al$ ($\be$).

It is important to note that, in deriving the translated \ac{OTPD}
functional expression in Eq.~\ref{EQ:TRE}, no dependence on
$\nabla\Pi(\br)$ is assumed. A scheme in which the \ac{OTPD} functional
depends explicitly upon $\nabla\Pi(\br)$ has also been
proposed.\cite{Sand2018}  The corresponding ``fully-translated'' (ft)
functionals are defined as
\begin{eqnarray}
\label{EQ:FTRE}
\EOTPD\left[\rho(\br), \Pi(\br), |\nabla\rho(\br)|,|\nabla\Pi(\br) \right|] \equiv \nonumber \\
\EXC[\tilde{\rho}_\al(\br), \tilde{\rho}_\be(\br), |\nabla\tilde{\rho}_\al(\br)|, |\nabla\tilde{\rho}_\be(\br)|]
\end{eqnarray}
Expressions for the fully-translated spin densities and their respective
gradients, taken from Ref. \citenum{Carlson2015c}, are provided in the
Appendix.

\section{Computational Details}
	\label{SEC:COMPDETAILS}

The 1- and \acp{2-RDM} entering Eqs.~\ref{EQ:RHO}-\ref{EQ:DPI} are
obtained from \ac{v2RDM-CASSCF} computations using the \ac{v2RDM-CASSCF}
plugin \cite{v2rdm_casscf} to the \textsc{Psi4} electronic structure
package. \cite{Turney:2012:1759} The v2RDM-CASSCF procedure is state specific;
the orbitals are optimized for the ground state of a given spin symmetry.  
For v2RDM-CASSCF-PDFT, we have
implemented translated and fully-translated versions of the \ac{SVWN3},
\cite{Gaspar1974,Slater1951,Vosko1980} \ac{PBE} \cite{Perdew1996a} and
\ac{BLYP} \cite{Becke1988,Lee1988} \ac{XC} functionals.  The \ac{XC}
energy, along with the one-electron and classical Coulomb contributions to
the MC-PDFT energy (Eq. \ref{EQ:EMCPDFT}) are evaluated using a new plugin
to \textsc{Psi4}.\cite{mcpdft}

The results of v2RDM-CASSCF-PDFT computations of the \acp{PEC} for N$_2$,
H$_2$O, and CN$^-$ are compared to those from reference computations
performed using \ac{CASPT2}, as implemented in the \textsc{Open-MOLCAS}
electronic structure package.  \cite{Aquilante2016} The standard imaginary
shift \cite{Forsberg1997} of 0.20 E$_{\rm h}$ and \textsc{Open-MOLCAS}'s default
value of 0.25 E$_{\rm h}$ for \ac{IPEA} \cite{Ghigo2004} were applied in all
\ac{CASPT2} computations.  All \ac{CASPT2} computations employed a
full-valence CI-driven CASSCF reference.


All \acs{v2RDM-CASSCF-PDFT} computations employ the \acl{DF} approximation
to the electron repulsion integrals.\cite{Whitten1973,Dunlap1979} The
N$_2$, H$_2$O, and CN$^-$ \acp{PEC} were computed using full-valence
\ac{v2RDM-CASSCF}, the cc-pVTZ basis set,\cite{Dunning:1989:1007} and the corresponding
JK-type auxiliary basis set.\cite{Weigend:2002:4285} The details of the full-valence active 
spaces can be found in the Supporting Information.  Singlet/triplet energy gaps for the linear
polyacene series were computed within the cc-pVTZ basis with the
corresponding JK-type auxiliary basis set.  Here, the v2RDM-CASSCF
computations employ a $(4k+2,4k+2)$ active space ($2k+1$ $\pi$ bonding
orbitals and $2k+1$ antibonding $\pi^*$ orbitals), where $k$ represents
the number of fused six-membered rings in the polyacene molecule.
Equilibrium geometries for the singlet and triplet states of the polyacene
series were determined at the v2RDM-CASSCF/cc-pVDZ level of theory using a
development version of the Q-Chem 5.1 electronic structure
package\cite{QChem4} and the analytic gradient implementation described in
Ref. \citenum{Mullinax:2018:JustAccepted}.

\section{Results and Discussion}
\label{SEC:RESULTS}

As evidence of the correctness of our implementation of v2RDM-CASSCF-PDFT,
we compare singlet/triplet energy splittings evaluated using this approach
to those obtained from conventional, CI-CASSCF-driven MC-PDFT. The test
set was comprised of the main-group divalent radicals considered in Ref.
\citenum{Bao2016}.  Equilibrium geometries for these molecules, as well as
MC-PDFT and experimentally-derived singlet/triplet energy splittings were
taken from that work and the references therein. The present
v2RDM-CASSCF-PDFT computations employed a full-valence active space, the
aug-cc-pVQZ basis set,\cite{Kendall:1992:6796} and the corresponding JK-type auxiliary basis set;
the active space details can be found in the Supporting Information.  When
enforcing the PQG and T2 $N$-representability conditions,
v2RDM-CASSCF-PDFT singlet/triplet gaps are in excellent agreement with
those from MC-PDFT (when using the same primary basis and active space). For
example, the mean absolute errors in the v2RDM-CASSCF-PDFT gaps, relative to
those from experiment, are 5.4, 7.1, 9.7, and 12.4 kcal mol$^{-1}$ when
using the translated PBE, translated BLYP, fully-translated PBE, and
fully-translated BLYP functionals, respectively.  The mean errors from
MC-PDFT are 5.5, 7.2, 9.6, and 12.3 kcal mol$^{-1}$ when using the same
functionals.  The agreement between v2RDM-CASSCF-PDFT and MC-PDFT is
slightly worse when v2RDM-CASSCF-PDFT employs RDMs that satisfy only the
PQG $N$-representability conditions; in this case, when using the same
functionals, v2RDM-CASSCF-PDFT displays mean errors of 4.9, 6.3, 8.5, and
10.9 kcal mol$^{-1}$.  Oddly enough, the gaps derived from computations
involving the PQG conditions are slightly more accurate than those from
v2RDM-CASSCF-PDFT with the PQG+T2 conditions or from MC-PDFT.  The
v2RDM-CASSCF-PDFT singlet/triplet energy spittings are tabulated alongside
those from (v2RDM-)CASSCF, MC-PDFT, and CASPT2 in the Supporting Information.

\subsection{Potential Energy Curves}\label{SUBSEC:PECs}
\begin{table*}[ht]
    \caption{Non-parallelity errors (mE$_{\rm h}$) in the potential
    energy curves relative to curves generated at the \ac{CASPT2}
    level of theory.} \label{TAB:ERR}
	\label{TAB:NPE}
\begin{tabular}{lcccccccc}
\hline
\hline

			Molecule$^a$ & $N$-representability  & v2RDM-CASSCF & tSVWN3 & tPBE & tBLYP & ftSVWN3 & ftPBE & ftBLYP\\[1pt]
			\hline
			\ce{N2}	 & \mr{3}{*}{PQG}    & 35.5         & 91.8   & 17.7 & 20.9  & 99.6    & 29.7  & 33.0  \\
			\ce{H2O} &                   & 55.5         & 64.0   & 24.5 & 23.7  & 66.4    & 27.2  & 28.6  \\
			\ce{CN-} &                   & 65.9         & 56.5   & 33.1 & 46.3  & 64.5    & 24.8  & 31.6  \\[4pt]
			\ce{N2}  & \mr{3}{*}{PQG+T2} & 29.5         & 92.6   & 18.3 & 18.2  & 100.8   & 30.1  & 33.5  \\
			\ce{H2O} &                   & 57.7         & 64.4   & 25.1 & 23.2  & 67.0    & 27.2  & 28.5  \\
			\ce{CN-} &                   & 68.0         & 63.6   & 31.3 & 43.8  & 71.8    & 23.4  & 29.6  \\[1pt]
\hline
\hline
		\end{tabular}
		\begin{tablenotes}
			\scriptsize

			\item $^a$ NPEs were evaluated for N--N, C--N, and
			O--O bond lengths in the ranges 0.7--5.0 \AA,
			0.7--5.0 \AA, and 0.6--5.0 \AA, respectively.

		\end{tablenotes}
%
\end{table*}

\begin{figure*}[!htpb]
	
	\caption{Potential energy curves for the dissocation of
		\ce{N2} within the cc-pVTZ basis set [(a), (c)], as well
        as their behavior in the limit of dissociation [(b), (d)].
		RDMs from v2RDM-CASSCF employed within v2RDM-CASSCF-PDFT satisfy
		the PQG $N$-representability conditions. Results are provided using both
		the translated [(a), (b)] and fully-translated [(c), (d)] 
		v2RDM-CASSCF-PDFT schemes.}
	
	\label{FIG:N2TZ}
	
	\includegraphics[width=0.8\linewidth, height=0.5\textheight]{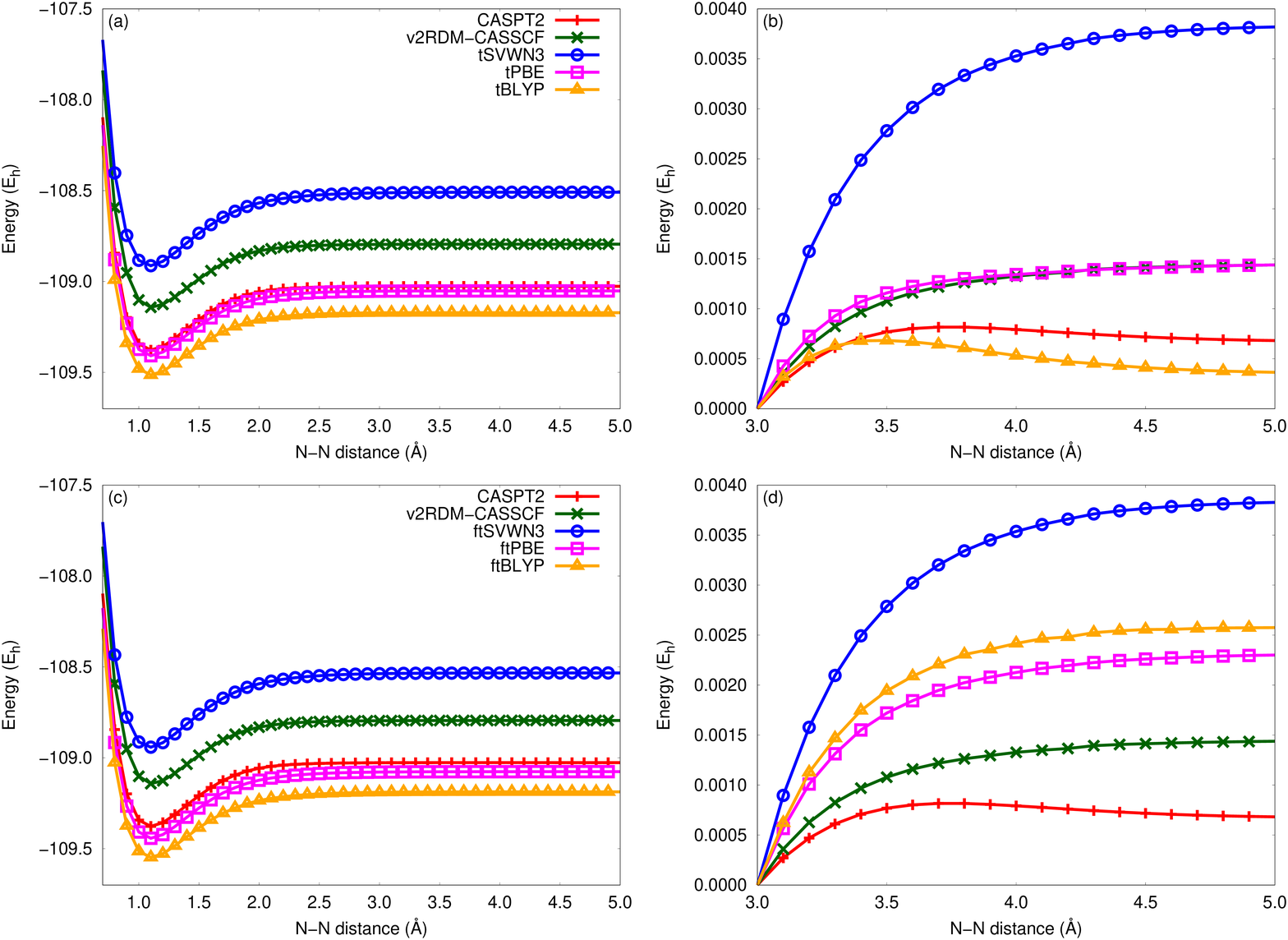}
	
\end{figure*}
Figure \ref{FIG:N2TZ}(a) provides dissociation curves for molecular
nitrogen computed at the v2RDM-CASSCF, CASPT2, and v2RDM-CASSCF-PDFT
levels of theory.  The 2-RDMs in the v2RDM-CASSCF computations satisfied
the PQG $N$-representability conditions. The v2RDM-CASSCF-PDFT
computations employed the translated variants of the SVWN3, PBE, and BLYP
functionals (denoted \acs{tSVWN3}, \acs{tPBE}, and \acs{tBLYP},
respectively).  Figure \ref{FIG:N2TZ}(b) illustrates the dissociation
limit for each method, where each curve is shifted such that the energy at
3.0 \AA~is zero E$_{\rm h}$.  The overall shape of the \ac{CASPT2}
\ac{PEC} is reasonably reproduced by both the \ac{v2RDM-CASSCF} and
v2RDM-CASSCF-PDFT methods, but we note that the dissociation limits of
CASPT2 (with an IPEA shift of 0.25 E$_{\rm h}$) and tBLYP show an
unphysical hump at 3.7 and 3.5 \AA, respectively.  Table \ref{TAB:NPE}
provides the \acp{NPE} for the PECs computed using each method.
The NPE is defined as the difference between
the maximum and minimum errors in each curve,
using \ac{CASPT2} as a reference.  
For N$_2$, the
\acp{NPE} in the v2RDM-CASSCF, tSWVN, \acs{tPBE}, and \acs{tBLYP} PECs are
35.5, 91.8, 17.7, and 20.9 mE$_{\rm h}$, respectively; for comparison, the
\ac{NPE} for CI-based CASSCF (not illustrated in Fig.  \ref{FIG:N2TZ}) is
30.1 mE$_{\rm h}$.  With the exception of v2RDM-CASSCF and tBLYP, the
maximum error contributing to the NPE occurs at 5.0 \AA, after CASPT2
begins to fail.  Table \ref{TAB:NPE} also provides \acp{NPE} computed when
the v2RDM-CASSCF and v2RDM-CASSCF-PDFT computations were carried out using
RDMs that satisfy the PQG+T2 $N$-representability conditions (the
corresponding dissociation curves can be found in the Supporting
Information).  The NPE is 29.5 mE$_{\rm h}$ for v2RDM-CASSCF, which is in
much better agreement with that from CI-based CASSCF.  The NPEs for
v2RDM-CASSCF-PDFT appear to be less sensitive to the $N$-representability
of the RDMs; the largest change we observe is for tBLYP, where the NPE is
reduced by 2.7 mE$_{\rm h}$.

The remaining two panels in Fig. \ref{FIG:N2TZ} illustrate the same PECs,
but the v2RDM-CASSCF-PDFT computations employed the fully-translated
variants of SVWN3, PBE, and BLYP functionals (denoted \acs{ftSVWN3},
\acs{ftPBE}, and ftBLYP, respectively).  In terms of absolute energies,
full translation universally lowers the energy obtained from
v2RDM-CASSCF-PDFT using all functionals, relative to the case in which
regular translation was employed.  The computed \acp{NPE} are considerably
worse, increasing by as much as 12.0 and 12.1 mE$_{\rm h}$ in the cases of
ftPBE and ftBLYP, respectively. Notably, full translation improves the
qualitative description of the dissociation limit in the case of ftBLYP.
Again, as can be seen in Table \ref{TAB:NPE}, the \acp{NPE} for
fully-translated functionals are quite insensitive to the
$N$-representability of the reference RDMs.

\begin{figure*}[!htpb]

    \caption{Potential energy curves for the symmetric dissocation of
    \ce{H2O} within the cc-pVTZ basis set [(a), (c)], as well
    as their behavior in the limit of dissociation [(b), (d)].
    RDMs from v2RDM-CASSCF employed within v2RDM-CASSCF-PDFT satisfy
    the PQG $N$-representability conditions. Results are provided using both
    the translated [(a), (b)] and fully-translated [(c), (d)] 
    v2RDM-CASSCF-PDFT schemes.}

    \label{FIG:H2OTZ}

    \includegraphics[width=0.8\linewidth, height=0.5\textheight]{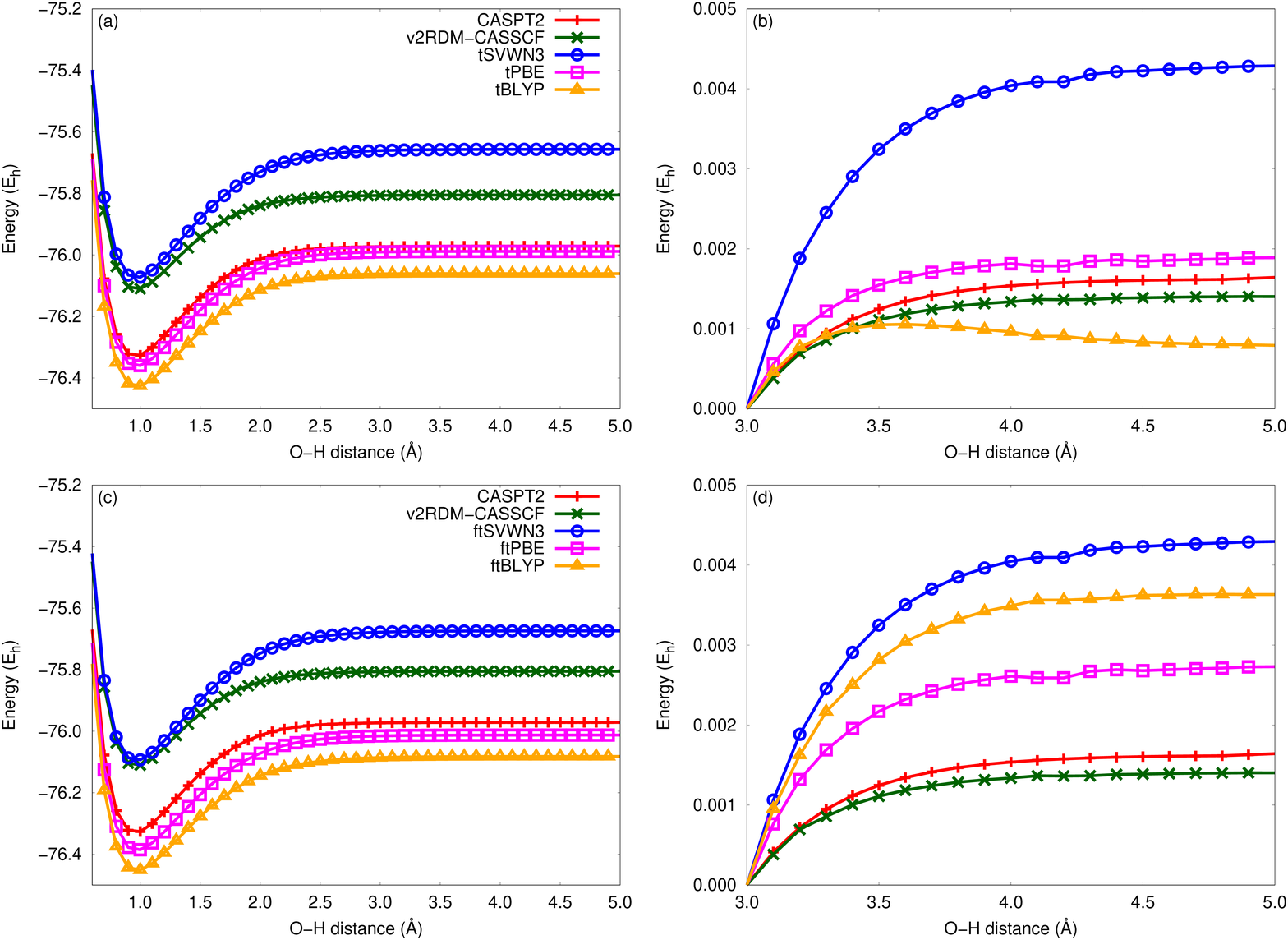}

\end{figure*}

Figure \ref{FIG:H2OTZ} illustrates PECs corresponding to the symmetric
double dissociation of H$_2$O, with a fixed H--O--H angle of
$104.5^\circ$, computed using the same levels of theory discussed above.
Similar conclusions can be drawn in this case, regarding the qualitative
agreement of the shapes of the PECs derived from v2RDM-CASSCF and
v2RDM-CASSCF-PDFT, relative to that from CASPT2.  However, tBLYP is the
only method that displays an unphysical hump in the dissociation limit.
The \acp{NPE} for v2RDM-CASSCF, \acs{tSVWN3}, \acs{tPBE}, and \acs{tBLYP}
are 55.5, 64.0, 24.5, and 23.7 mE$_{\rm h}$, respectively, when using RDMs
that satisfy the PQG conditions.  As noted above, the \ac{NPE} for
v2RDM-CASSCF is in much better agreement with that from CI-based CASSCF
(both are 57.7 mE$_{\rm h}$) when the RDMs satisfy the T2 condition.  Again, full
translation universally lowers the energy obtained from v2RDM-CASSCF-PDFT,
and the corresponding \acp{NPE} are insensitive to the
$N$-representability of the underlying RDMs. The largest change observed in the
NPE is an increase of 0.6 mE$_{\rm h}$, in the case of \acs{ftSVWN3} and \acs{tPBE}.


As an additional measure of the quality of the computed PECs, we consider
the dissociation energy for N$_2$ and the double dissociation energy for
H$_2$O. These values, along with estimates of the equilibrium bond lengths
in these molecules, are presented in Table \ref{TAB:DE};  errors in
computed dissociation energies are provided in parentheses.  Here, an 
approximation to the dissociation energy is defined as the difference between energy computed
at the equilibrium geometry and at stretched geometries with N--N or O--H
bond lengths of 5.0 \AA~(the H--O--H angle of H$_2$O was fixed at 104.5$^{\circ}$).
The equilibrium bond lengths were estimated by fitting a
quadratic function to each PECs around equilibrium and identifying the minimum
in that function.  The worst estimates of the dissociation energy are
provided by v2RDM-CASSCF and either translated or fully-translated SVWN
PDFT functionals, regardless of the $N$-representability conditions
imposed in the reference v2RDM-CASSCF computations.  Significanlty
improved results are obtained from the tPBE, ftPBE, and ftBLYP
functionals; for both molecules, these functionals yield dissociation
energies that are accurate to a few kcal mol$^{-1}$.  The performance of
the tBLYP functional is much worse for N$_2$ than it is for H$_2$O, which
is a consequence of the unphysical hump in the tPBE PEC for N$_2$ that was
discussed above.  As with the NPE, dissociation energies from
v2RDM-CASSCF-PDFT are less sensitive than those from v2RDM-CASSCF to the
$N$-representability of the RDMs. For N$_2$, the v2RDM-CASSCF dissociation
energies change by approximately 6 kcal mol$^{-1}$ depending on whether or
not the T2 condition is enforced; on the other hand, the corresponding
v2RDM-CASSCF-PDFT dissociation energies differ by roughly 2 kcal
mol$^{-1}$.  For H$_2$O, the v2RDM-CASSCF double dissociation energy
changes by about 1 kcal mol$^{-1}$ when enforcing the T2 condition, while
the corresponding v2RDM-CASSCF-PDFT values change by only 0.1--0.2 kcal
mol$^{-1}$.

\begin{table*}[ht]

    \caption{Equilibrium bond lengths ($R$, \AA) and dissociation energies
    ($D_e$, kcal mol${-1}$) obtained from experiment, as well as from
    v2RDM-CASSCF and various \ac{PDFT} functionals within the \ac{cc-pVTZ}
    basis.  Errors in computed dissociation energies are provided in
    parentheses.}

    \label{TAB:DE}
    \begin{tabular}{lccccc}
    	\hline\hline
    	\mr{2}{*}{Method} & \mr{2}{*}{$N$-representability} & \mc{2}{c}{\ce{N2}} & \mc{2}{c}{\ce{H2O}}   \\ \cline{3-4}\cline{5-6}
    	                  &                                 &        $R$         &        $D_e$        &      $R$      &     $D_e$      \\[1pt] \hline
    	v2RDM-CASSCF      &         \mr{7}{*}{PQG}          &        1.12        &    217.8 (-10.7)    &     0.972     & 192.5 (-39.7)  \\
    	tSVWN3            &                                 &        1.11        &    253.6 (25.1 )    &     0.973     & 262.2 (30.0 )  \\
    	tPBE              &                                 &        1.11        &    223.5 (-5.1 )    &     0.975     & 233.9 (1.7  )  \\
    	tBLYP             &                                 &        1.11        &    214.4 (-14.1)    &     0.976     & 230.2 (-2.0 )  \\
    	ftSVWN3           &                                 &        1.10        &    256.0 (27.5 )    &     0.971     & 264.5 (32.3 )  \\
    	ftPBE             &                                 &        1.11        &    229.7 (1.1  )    &     0.976     & 234.9 (2.7  )  \\
    	ftBLYP            &                                 &        1.11        &    226.0 (-2.5 )    &     0.978     & 232.6 (0.4  )  \\[3pt]
    	v2RDM-CASSCF      &        \mr{7}{*}{PQG+T2}        &        1.12        &    212.0 (-16.5)    &     0.971     & 191.5 (-40.7)  \\
    	tSVWN3            &                                 &        1.11        &    255.9 (27.4 )    &     0.974     & 262.2 (30.1 )  \\
    	tPBE              &                                 &        1.11        &    225.5 (-3.0 )    &     0.976     & 233.8 (1.6  )  \\
    	tBLYP             &                                 &        1.11        &    216.6 (-11.9)    &     0.977     & 230.1 (-2.1 )  \\
    	ftSVWN3           &                                 &        1.10        &    258.5 (29.9 )    &     0.972     & 264.6 (32.4 )  \\
    	ftPBE             &                                 &        1.11        &    231.2 (2.7  )    &     0.976     & 234.7 (2.5  )  \\
    	ftBLYP            &                                 &        1.11        &    227.8 (-0.7 )    &     0.978     & 232.4 (0.3  )  \\ \hline
    	Experiment        &                                 &   1.098$^{a,c}$    &    227.8$^{a,c}$    & 0.958$^{b,c}$ &  232.2$^{d}$   \\ \hline\hline
    \end{tabular}
    \begin{tablenotes}
        \footnotesize
        \item $^a$ From Refs. \citenum{HerzbergBook4} and \citenum{Ermler:1982:1305}
        \item $^b$ From Refs. \citenum{Darwent:1970:31} and \citenum{Hoy:1979:1}
        \item $^c$ From Ref. \citenum{CCCBDB}
        \item $^d$ From Ref. \citenum{Kurth:1999:889}
    \end{tablenotes}
\end{table*}

\begin{figure*}[!htpb]

    \caption{Potential energy curves for the dissocation of
	\ce{CN-} within the cc-pVTZ basis set [(a), (c)], as well
    as their behavior in the limit of dissociation [(b), (d)].  RDMs from v2RDM-CASSCF
	employed within v2RDM-CASSCF-PDFT satisfy the PQG
	$N$-representability conditions. Results are provided using both
	the translated [(a), (b)] and fully-translated [(c), (d)]
	v2RDM-CASSCF-PDFT schemes.}

	\label{FIG:CNTZ}

    \includegraphics[width=0.8\linewidth, height=0.5\textheight]{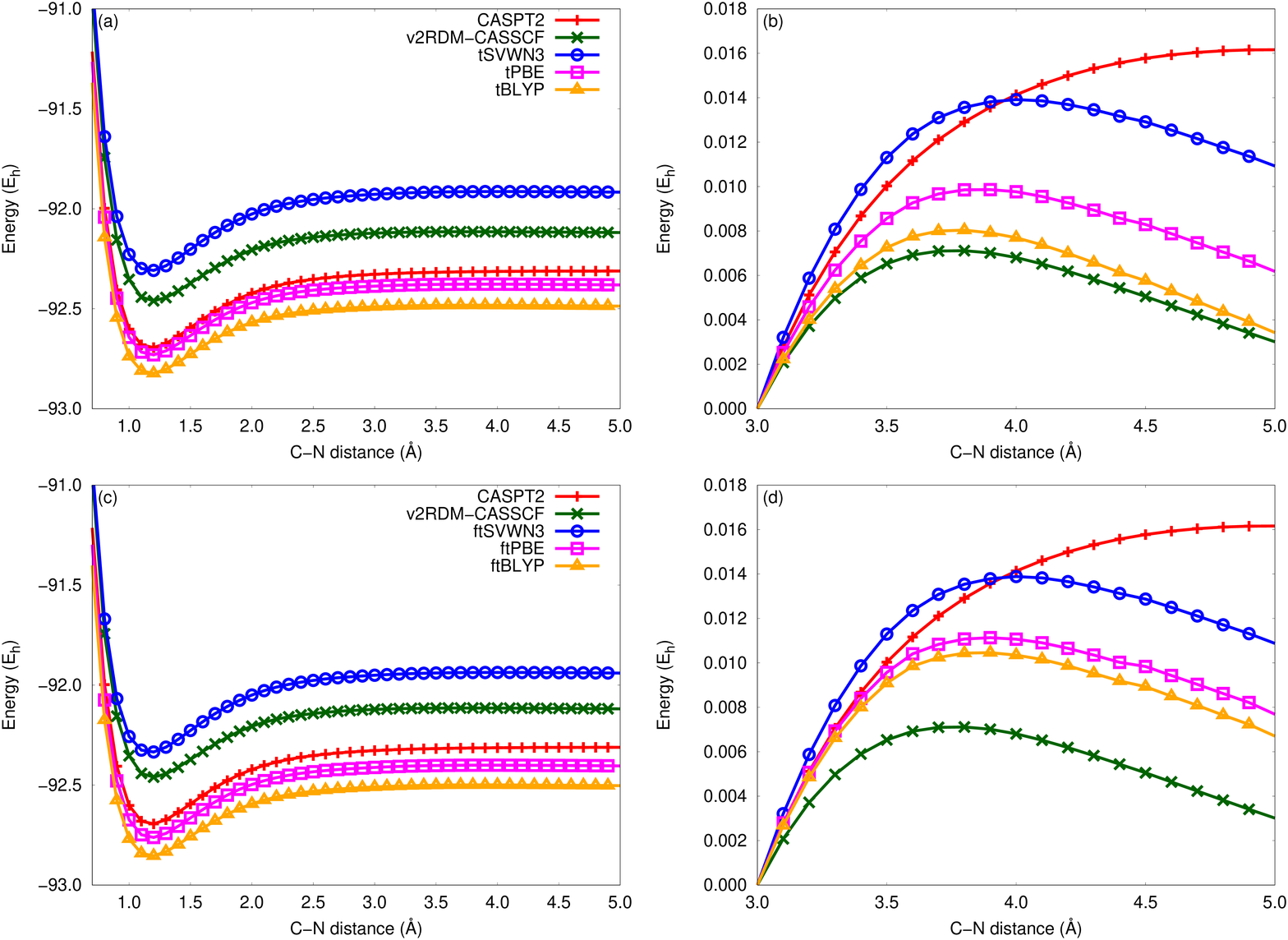}

\end{figure*}

The last PECs considered are those for the dissociation of \ce{CN-} in the
cc-pVTZ basis set, which are depicted in Fig. \ref{FIG:CNTZ}.  Some of the
qualitative features of the v2RDM-CASSCF and v2RDM-CASSCF-PDFT derived
PECs follow the same trends observed for N$_2$ and H$_2$O; for
example, fully-translated functionals yield lower energies than translated
ones, and tPBE and ftPBE are the most accurate flavors of
v2RDM-CASSCF-PDFT, as measured by the absolute deviations from CASPT2.  On
the other hand, one notable difference stands out in this case: all
v2RDM-based methods behave qualitatively incorrectly in the dissociation
limit.  It is well known that v2RDM-based approaches dissociate
heteronuclear diatomic molecules into fractionally charged
species;\cite{VanAggelen:2009:5558} the description of CN$^-$ with
v2RDM-CASSCF is one such case.  This issue stems from a lack of derivative
discontinuity in the energy as a function of electron number in isolated
atoms, which has long been known to impact the quality of the description
of the dissociation limit.  \cite{PerdewPRL1982}  It is not surprising
that v2RDM-CASSCF-PDFT built upon reference RDMs from v2RDM-CASSCF would
display the same incorrect behavior.  We also note that enforcing partial
three-particle $N$-representability does not improve the situation;
additional dissociation curves that demonstrate this failure can be found
the Supporting Information. 


\subsection{Singlet/Triplet Energy Gaps in Polyacene Molecules}\label{SUBSEC:STGAPS}

The electronic structure of the linear polyacene series has long been of
interest to experimentalists and theorticians alike.  The optical
properties of these molecules, particularly their propensity to undergo
singlet
fission,\cite{Tayebjee:2015:2367,Hanna:2006:074510,Zimmerman:2011:19944,Wilson:2013:1330,Izadnia:2017:2068}
make them desirable components in photovoltaic
devices.\cite{Yoo:2004:5427,Mayer:2004:6272,Chu:2005:243506,Paci:2006:16546,Rand:2007:659,
Shao:2007:103501} The instability of the longer members of the series
usually limits practical devices to those containing four or five fused
benzene rings, but synthesis of polyacenes with up to nine fused benzene
rings has been
reported.\cite{birksphotophysics,schiedt1997photodetachment,Angliker:1982:208,sabbatini1982quenching,burgos1977heterofission,Mondal:2009:14281,Tonshoff:2010:4125,Zade:2010:4012}

The fascinating electronic structure of the larger members of the
polyacene series has fueled a series of contentious interpretations of the
results of state-of-the-art electronic structure computations.  These
controversies began with debates over the ground spin state of the longer
members of the
series\cite{Houk:2001:5517,Bendikov:2004:7416,Jiang:2008:332,Hajgato:2009:224321,Wu:2015:2003,Ibeji:2015:9849}
and have evolved into a discussion over the degree to which the
lowest-energy singlet state can be considered a closed-shell di- or
polyradical.\cite{Bendikov:2004:7416,Hachmann:2007:134309,Hajgato:2009:224321,Gidofalvi2008,Mizukami:2013:401,FossoTande:2016:2260,FossoTande:2016:423,Schriber:2016:161106,Lee:2017:602,Battaglia:2017:3746,Lehtola:2018:547,Dupuy:2018:134112}
The former question has been settled; it is generally agreed upon that the
singlet state is lower in energy than the triplet state for all linear
acene molecules.  Only recently, however, has it become clear that even
methods capable of describing non-dynamical correlation effects in large
active spaces (e.g.  DMRG- or v2RDM-based CASSCF) tend to overestimate the
polyradical character of the larger members of the series when correlations
among the $\sigma/\sigma*$ network are
ignored.\cite{Lee:2017:602,Battaglia:2017:3746,Lehtola:2018:547,Dupuy:2018:134112}
A detailed history of the progression of these controversies is recounted
in Ref. \citenum{Dupuy:2018:134112}.

In this Section, we explore the utility of v2RDM-CASSCF-PDFT for modeling
the singlet/triplet energy gap in linear acene molecules.  The literature
is cluttered with estimates of this quantity generated using a variety of
MR methods, including v2RDM-,
\cite{FossoTande:2016:2260,FossoTande:2016:423,Pelzer:2011:5632} DMRG-,
\cite{Hachmann:2007:134309,Mizukami:2013:401,Sharma:2018:arXiv} adaptive
CI (ACI)-, \cite{Schriber:2016:161106,Schriber:2018:arXiv} and
MC-PDFT-based\cite{Ghosh:2017:2741,Sharma:2018:arXiv} approaches, as well
as with \ac{MR-AQCC}\cite{Plasser:2013:2581,Horn:2014:1511} and quantum
Monte-Carlo methods.\cite{Dupuy:2018:134112}.  Nevertheless, no one
approach has emerged as a panacea for this particular problem, for a
variety of reasons.  First, as mentioned above, nondynamical correlation
effects are quite important for large members of the series, and, yet,
even active spaces comprised of the full $\pi/\pi^*$ valence space fail to
correctly describe the onset of closed-shell diradical behavior.  A proper
description of these systems requires that one at least consider dynamical
correlation effects, if not additional nondynamical correlation effects
among the remaining valence orbitals.  Second, most studies employ
inconsistent levels of theory to evaluate the equilibrium molecular
geometries and the singlet/triplet energy gaps; equilibrium geometries are
usually determined using restricted or unrestricted DFT and the B3LYP
functional.  Such a choice often results in singlet/triplet energy gap
curves that are not completely
smooth.\cite{Ghosh:2017:2741,FossoTande:2016:2260,Dupuy:2018:134112}

\begin{figure}[!htpb]
	\includegraphics[]{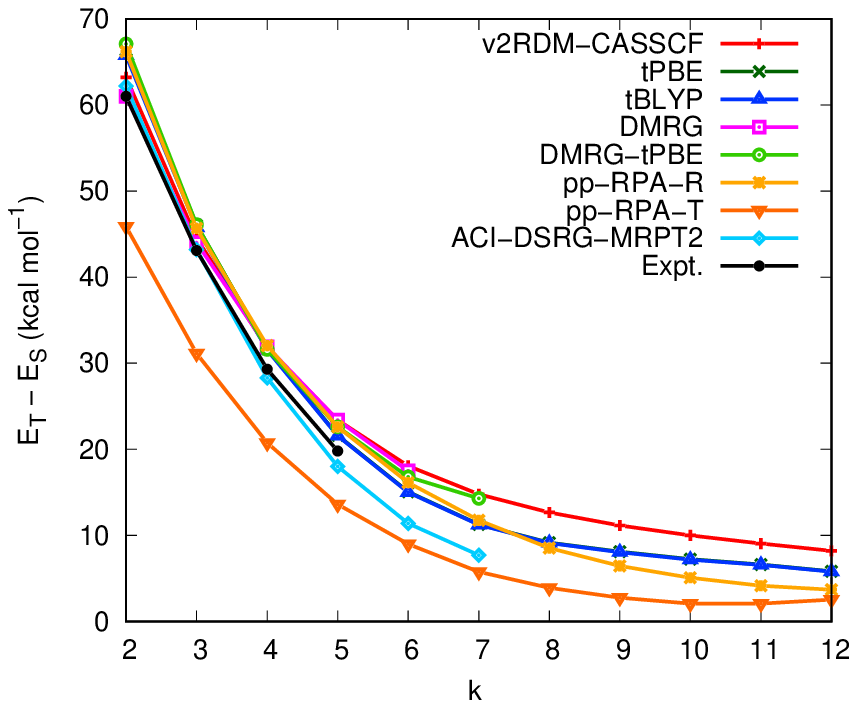}
	\caption{Singlet-triplet energy gaps of the linear polyacene
	series.  The label ``$k$'' refers to the number of fused benzene
	rings in each molecule.}
	\label{FIG:ACENES}
\end{figure}

Figure \ref{FIG:ACENES} illustrates the singlet/triplet energy gap for the
linear polyacene series computed using a variety of methods, including
v2RDM-CASSCF and v2RDM-CASSCF-PDFT (labeled tPBE and tBLYP).  All
v2RDM-based computations employed an active space comprised of the
$\pi/\pi*$ valence space (4$k$+2 electrons in 4$k$+2 orbitals, where $k$
is the number of fused benzene rings in the molecule).  The v2RDM-based
energy gaps presented here were computed within the cc-pVTZ basis set
using equilibrium geometries optimized for the singlet and triplet states
at the v2RDM-CASSCF/cc-pVDZ level of theory.
We thus have some guarantee
that the equilibrium geometries reflect the presence of nondynamical
correlation effects, but we note that the structures display signatures of
the overestimation of polyradical character mentioned above (see Refs.
\citenum{Dupuy:2018:134112} and \citenum{Mullinax:2018:JustAccepted} for a
discussion of these effects).  Nonetheless, the v2RDM-CASSCF energy gaps
agree well with those from DMRG\cite{Hachmann:2007:134309}, despite the
fact that the DMRG data were generated using a smaller basis set and
B3LYP-derived geometries.  As expected, the dynamical correlation effects
captured by v2RDM-CASSCF-PDFT close the singlet/triplet energy gap
considerably for larger molecules, and we note that the tPBE and tBLYP
functionals yield essentially equivalent gaps for all molecules.
Surprisingly, DMRG-PDFT and v2RDM-CASSCF-PDFT give quite different results
for hexacene and heptacene.  The DMRG-PDFT data of Ref.
\citenum{Sharma:2018:arXiv} were generated within the 6-31+G(d,p) basis
set, using B3LYP-derived geometries, but, these differences are still
unanticipated, considering the good agreement we observe between the DMRG
results of Ref. \citenum{Hachmann:2007:134309} and the present
v2RDM-CASSCF results.

Figure \ref{FIG:ACENES} also includes data generated using a combined ACI
/ second-order perturbative multireference driven similarity
renormalization group (DSRG-MRPT2) approach and the particle-particle
random phase approximation (pp-RPA), taken from Refs.
\citenum{Schriber:2018:arXiv} and \citenum{Yang:2016:E5098}, respectively.
Of all methods considered here, ACI-DSRG-MRPT2 yields the best agreement
with experimental results, up to pentacene.  Assuming that ACI and
DSRG-MRPT2 can be extended to larger members of this series,
ACI-DSRG-MRPT2 appears to be an extrememly promising approach for this
problem.  Direct comparisons to pp-RPA results are complicated by the fact
that these gaps are vertical, whereas the present results and all others
reproduced here are adiabatic.  The label pp-RPA-R (pp-RPA-T) refers to
computations performed at equilibrium geometries generated for the singlet
(triplet) at the restricted (unrestricted) B3LYP level of theory.  The
pp-RPA data were also generated in a smaller basis set (cc-pVDZ) than was
used in the present v2RDM-CASSCF-PDFT computations.  The pp-RPA-T predicts
singlet/triplet energy gaps that are significantly lower than those
predicted by all other methods. Further, pp-RPA-T is the only method that
yields gaps that do not decrease monotonically with increasing polyacene
length.  We note that the gaps from pp-RPA-R, however, are in good
agreement with the present tPBE/tBLYP gaps, for acene molecules smaller
than nonacene.

Lastly, we consider the singlet/triplet energy gap extrapolated to the
limit of an infinitely large molecule.  The computed singlet/triplet
energy gaps ($\EST$) for finite molecules were fit to an exponential decay
formula of the form
\begin{equation}\label{EQ:FIT}
	\EST(k) = a e^{-k/b} + c 
\end{equation}
where, $a$, $b$ and $c$ are adjustable parameters, and $k$ represents the
number of fused benzene rings in the molecule. In the limit that $k$
approaches infinity, $c\sim\EST(\infty)$.  Table \ref{TAB:FIT} summarizes
the fitting parameters and predictions for $\EST(\infty)$ for the subset of
methods considered above for which data are available up to $k=12$.  The
largest predicted gap is obtained by v2RDM-CASSCF, which indicates that
the limited considerations of nondynamical correlation, combined with a
lack of dynamical correlation, artificially stabilize the singlet state.
The gaps predicted by tPBE and tBLYP are 4.87 and 4.79 kcal mol$^{-1}$,
respectively.  These values are in good agreement with the ``best
estimate'' value of 5.06 kcal mol$^{-1}$ of Ref.
\citenum{Ibeji:2015:9849}, which was computed using a combination of
spin-flip coupled-cluster and spin-flip time-dependent DFT.  The smallest
estimates for the infinite-acene singlet/triplet gap are given by pp-RPA-R
and pp-RPA-T.  



\begin{table}[h]
	\setlength{\tabcolsep}{7pt}
	\center
	\caption{Fitting formulas for singlet-triplet energy gaps of polyacenes as a function of number of fused benzene rings $k$ where $k \in [2,12]$} \label{TAB:FIT}
	\begin{tabular}{lc}
		\hline\hline
		Level of Theory                      &        Fitting Formula        \\ \hline
		\acs{v2RDM-CASSCF}/\acs{cc-pVTZ}$^a$ & $127.79 \exp(-k/2.39) + 7.85$ \\
		\acs{tPBE}/\acs{cc-pVTZ}$^a$         & $147.15 \exp(-k/2.32) + 4.87$ \\
		\acs{tBLYP}/\acs{cc-pVTZ}$^a$        & $145.04 \exp(-k/2.33) + 4.79$ \\
		\acs{pp-RPA}-R/\acs{cc-pVDZ}$^b$     & $137.04 \exp(-k/2.63) + 2.11$ \\
		\acs{pp-RPA}-T/\acs{cc-pVDZ}$^b$     & $105.87 \exp(-k/2.37) + 0.81$ \\ \hline\hline
	\end{tabular}
	\begin{tablenotes}
		\scriptsize
		\item $^a$ Geometries are optimized at v2RDM-CASSCF/cc-pVDZ level of theory.
		\item $^b$ Geometries are optimized at B3LYP/6--31G(d) level of theory.
	\end{tablenotes}
\end{table}

\section{Conclusions}
\label{SEC:CONCLUSIONS}

Multiconfigurational pair-density functional theory provides a
conceptually and technically straightfoward framework within which one can
combine the reliable description of nondynamical correlation effects
afforded by multireference methods with the simplicity of DFT for modeling
dynamical correlation.  In practice, the computational cost of MC-PDFT is
dominated entirely by the effort required to generate the 1- and 2-RDM
using the underlying MR approach.  Hence, polynomially-scaling approaches
to the nondynamical correlation problem (e.g. v2RDM-CASSCF,
DMRG-CASSCF\cite{Sharma:2018:arXiv}, or pair coupled-cluster
doubles\cite{Garza2015}) are naturally suited to this purpose.
Accordingly, we have presented an implementation of v2RDM-CASSCF-PDFT and
benchmarked its performance on challenging MR problems.

We applied v2RDM-CASSCF-PDFT with the translated and fully-translated
variants of the PBE, BLYP, and SVWN3 functionals to the dissociation of
N$_2$ and CN$^{-}$ and to the double dissociation of H$_2$O.  In general,
the best absolute agreement with potential energy curves generated at the
CASPT2 level of theory was obtained using tPBE.  It is notable that the
quality of the v2RDM-CASSCF-PDFT curves, as measured by the
non-parallelity error relative to CASPT2, was somewhat insensitive to the
$N$-representability of the underlying 2-RDM.  A similar insensitivity was
observed in the dissociation energy for N$_2$ and
the double dissociation energy for H$_2$O computed using v2RDM-CASSCF-PDFT.  These results are potentially
important because the cost of imposing three-particle $N$-representability
conditions can be prohibitive for large systems.  It seems that additional
$N$-representability conditions may be necessary insofar as they improve
the quality of the 1-RDM and on-top pair-density entering the
v2RDM-CASSCF-PDFT energy expression. Further, we should note that even an
exact treatment of the $N$-representability problem will not actually
guarantee any improvement in the v2RDM-CASSCF-PDFT energy. Indeed, Sharma
{\em et al.} have shown\cite{Sharma:2018:660} for H$_2$ that the MC-PDFT
energy does not converge to the exact (full CI) result even when the exact
on-top pair-density is available.  Regardless, additional studies are
necessary to fully explore the role of $N$-representability in generating
accurate densities and on-top pair densities and to determine whether the
trends we have observed are transferable to other systems.

We also applied the tPBE and tBLYP functionals to the singlet/triplet
energy gap of the linear polyacene series; these functionals predict the
gap in the limit of an infinitely long acene molecule to be 4.87 and 4.79
kcal mol$^{-1}$, respectively.  We note that a similar study has been
carried out using MC-PDFT where the 1- and 2-RDM were generated using the
generalized active space self-consistent field (GASSCF)
method.\cite{Ghosh:2017:2741} However, direct comparisons to the data
presented in Ref.  \citenum{Ghosh:2017:2741} are difficult because the
GASSCF and GASSCF-PDFT results are sensitive to the partitioning chosen
for the active space. For example, one choice leads to the prediction that
the GASSCF-PDFT-derived singlet/triplet gap in the large molecule limit
closes, relative to that from GASSCF, while another choice leads to the
prediction that it opens.  Further, regardless of how the active space was
partitioned in that work, GASSCF-PDFT failed to yield a smooth
singlet/triplet energy gap curve.



Lastly, we note a practical similarity between MC-PDFT and the density
corrected DFT described by Burke and
coworkers.\cite{Kim:2013:073003,Kim:2014:18A528,Kim:2015:3802}  The latter
approach addresses so-called ``density-driven'' errors by evaluating the
exchange-correlation energy with an ``accurate'' density that differs from
the self-consistent density one would normally obtain with a given
functional.  MC-PDFT can be viewed as a generalization of
density-corrected DFT where the density used to evaluate the exchange
correlation energy is obtained by translation\cite{LiManni2014} of the
density generated by an MR method.  As demonstrated in Ref.
\citenum{Garza2015}, the MR method need not be CASSCF-based; it could be any
method that produces a good 1-RDM and on-top pair density.  The
v2RDM-CASSCF-PDFT PECs for CN$^-$ highlight the fact that the approach is
only as reliable as the densities generated by the MR method.

\vspace{0.5cm}
\section{Appendix}
The fully-translated densities and gradients entering the fully-translated
\ac{OTPD} functional (defined in Eq. \ref{EQ:FTRE}) are given by
\cite{Carlson2015c}
\footnotesize
\begin{align}\label{EQ:FTR}
\tilde{\rho}_\sig(\br) &=\begin{cases}
\frac{\rho(\br)}{2} \left(1 + c_\sig \zetr(\br)\right)  &       \quad R(\br) < R_0                              \\[8pt]
\frac{\rho(\br)}{2} \left(1 + c_\sig \zeftr(\br)\right) &       \quad R_0 \leq R(\br) \leq R_1          \\[8pt]
\frac{\rho(\br)}{2}             &       \quad R(\br) > R_1
\end{cases}             \qquad
\end{align}
and
\begin{align}
&\nabla\tilde{\rho}_\sig(\br) &=\begin{cases}
\frac{\nabla\rho(\br)}{2} \left(1 + c_\sig \zetr(\br)\right)  + c_\sig \frac{\rho(\br)}{2} \nabla\zetr(\br)         &   \quad R(\br) < R_0                              \\[8pt]
\frac{\nabla\rho(\br)}{2} \left(1 + c_\sig \zeftr(\br)\right) + c_\sig \frac{\rho(\br)}{2} \nabla\zeftr(\br)    &       \quad R_0 \leq R(\br) \leq R_1  \\[8pt]
\frac{\nabla\rho(\br)}{2}               &       \quad R(\br) > R_1
\end{cases}     \\ \nonumber
\end{align}
\normalsize
where $R_0 = 0.9$ and $R_1 = 1.15$. \cite{Carlson2015c, Sand2018} The
fully-translated spin-polarization factor $\zeftr(\br)$ is taken to be
\begin{equation}\label{EQ:ZFTR}
        \zeftr = A \De R^5(\br) + B \De R^4(\br) + C \De R^3(\br)
\end{equation}
where,  $\De R(\br) = R(\br) - R_1      $ and \cite{Carlson2015c, Sand2018}
\begin{align}\label{EQ:FTRPARAMS}
        A &= -475.60656009      \\
        B &= -379.47331922      \\
        C &= -85.38149682       
\end{align}
The gradients of the translated and fully-translated spin-polarization
factors are \cite{Carlson2015c, Sand2018}
\small
\begin{gather}
        \nabla\zetr(\br)  = -\frac{1}{2} \frac{\nabla R(\br)}{\zetr(\br)}               \\[4pt]
        \nabla\zeftr(\br) = \nabla R(\br)~ [ 5 A \De R^4(\br) + 4 B \De R^3(\br) + 3 C \De R^2(\br) ]
\end{gather}
\normalsize
where the gradient of the on-top ratio is \cite{Sand2018}
\begin{equation}
        \nabla R(\br) = \frac{4 \nabla\Pi(\br)}{\rho^2(\br)} - \frac{8 \Pi(\br) \nabla\rho(\br)}{\rho^3(\br)}.
\end{equation}

\vspace{0.5cm}
{\bf Supporting information.}  
Active space specifications, singlet/triplet energy gaps for
non-conjugated main-group divalent radicals, potential energy curves for
N$_2$, CN$^{-}$ and H$_2$O dissociation computed using v2RDM-CASSCF /
v2RDM-CASSCF-PDFT and the PQG+T2 $N$-representability conditions, and
singlet/triplet energy gaps for linear polyacene molecules computed using
v2RDM-CASSCF / v2RDM-CASSCF-PDFT and RDMs that satisfy the PQG
$N$-representability conditions.

\vspace{0.5cm}
{\bf Acknowledgments} This material is based upon work supported by the
Army Research Office Small Business Technology Transfer (STTR) program
under Grant No.  W911NF-16-C-0124.


\label{SEC:ACRONYMS}
\begin{acronym}
	\acro{SI}{Supporting Information}
	\acro{RASSCF}{restricted active-space \acl{SCF}}
	\acro{GASSCF}{generalized active-space \acl{SCF}}
	\acro{CASSCF}{\acl{CAS} \acl{SCF}}\
	\acro{ORMAS}{occupation-restricted multiple active-space}
	\acro{DE}{delocalization error}
	\acro{LE}{localization error}
	\acro{SIE}{self-interaction error}
	\acro{CC}{coupled-cluster}
	\acro{MO}{molecular orbital}
	\acro{BO}{Born-Oppenheimer}
	\acro{CPO}{correlated participating orbitals}
	\acro{SCF}{self-consistent field}
	\acro{CAS}{complete active space}
	\acro{PT}{perturbation theory}
	\acro{CASPT2}{\acl{CAS} second-order \acl{PT}}
	\acro{IPEA}{ionization potential electron affinity}
	\acro{MC}{multiconfiguration}
	\acro{HF}{Hartree-Fock}
	\acro{MR}{multireference}
	\acro{MR-AQCC}{\acl{MR}-averaged quadratic \acl{CC}}
	\acro{CI}{configuration interaction}
	\acro{ACI}{adaptive \acl{CI}}
	\acro{FCI}{full \acl{CI}}
	\acro{ACSE}{anti-Hermitian contracted Schr\"odinger equation}
	\acro{DSRG}{driven similarity renormalization group}
	\acro{DMRG}{density matrix renormalization group}
	\acro{RPA}{random-phase approximation}
	\acro{pp-RPA}{particle-particle \acl{RPA}}
	\acro{CSE}{contracted Schr\"odinger equation}
	\acro{ACSE}{anti-Hermitian \acl{CSE}}
	\acro{aug-cc-pVQZ}{augmented correlation-consistent polarized-valence quadruple-$\ze$}
	\acro{cc-pVTZ}{correlation-consistent polarized-valence triple-$\ze$}
	\acro{cc-pVDZ}{correlation-consistent polarized-valence double-$\ze$}
	\acro{WFT}{wave function theory}
	\acro{pCCD}{pair coupled-cluster doubles}
	\acro{DF}{density-fitting}
	\acro{ERI}{electron-repulsion integral}
	\acro{ZPVE}{zero-point vibrational energy}
	\acro{MCSCF}{\acl{MC} \acl{SCF}}
	\acro{DFT}{density functional theory}
	\acro{PDFT}{pair-\acl{DFT}}
	\acro{OTPD}{on-top pair-density}
	\acro{MC-PDFT}{\acl{MC}-\acl{PDFT}}
	\acro{ft}{full translation}
	\acro{tr}{conventional translation}
	\acro{CASSCF}{\acl{CAS} \acl{SCF}}
	\acro{RDM}{reduced-density matrix}
	\acro{RDMs}{reduced-density matrices}
	\acro{1-RDM}{one-electron \acl{RDM}}
	\acro{2-RDM}{two-electron \acl{RDM}}
	\acro{3-RDM}{three-electron \acl{RDM}}
	\acro{4-RDM}{four-electron \acl{RDM}}
	\acro{HRDM}{hole \acl{RDM}}
	\acro{FP-1}{frontier partition with one set of interspace excitations}
	\acro{SF}{spin-flip}
	\acro{CCSD}{coupled-cluster with singles and doubles}
	\acro{SF-CCSD}{\acl{SF}-\acl{CCSD}}
	\acro{1-HRDM}{one-hole \acl{RDM}}
	\acro{2-HRDM}{two-hole \acl{RDM}}
	\acro{PEC}{potential energy curve}
	\acro{CCSDT}{coupled-cluster, singles doubles and triples}
	\acro{KS}{Kohn-Sham}
	\acro{HK}{Hohenberg-Kohn}
	\acro{XC}{exchange-correlation}
	\acro{LSDA}{local spin-density approximation}
	\acro{GGA}{generalized gradient approximation}
	\acro{MP2}{second-order M\o ller-Plesset \acl{PT}}
	\acro{SVWN3}{Slater and Vosko-Wilk-Nusair random-phase approximation expression III}
	\acro{PBE}{Perdew-Burke-Ernzerhof}
	\acro{LYP}{Lee-Yang-Parr}
	\acro{BLYP}{Becke and \acl{LYP}}
	\acro{B3LYP}{Becke-3-\acl{LYP}}
	\acro{tSVWN3}{translated \acl{SVWN3}}
	\acro{tPBE}{translated \acl{PBE}}
	\acro{tBLYP}{translated \acl{BLYP}}
	\acro{ftSVWN3}{fully \acl{tSVWN3}}
	\acro{ftPBE}{fully \acl{tPBE}}
	\acro{ftBLYP}{fully \acl{tBLYP}}
	\acro{v2RDM}{variational \acs{2-RDM}}
	\acro{v2RDM-CASSCF}{\acl{v2RDM}-driven \acs{CASSCF}}
	\acro{v2RDM-CAS}{\acl{v2RDM}-driven \acl{CAS}}
	\acro{CAS-PDFT}{\acl{CAS} \acl{PDFT}}
	\acro{v2RDM-CASSCF-PDFT}{\acl{v2RDM} \acl{CASSCF} \acl{PDFT}}
	\acro{MAX}{maximum absolute error}
	\acro{CAM}{Coulomb-attenuating method}
	\acro{MAE}{mean absolute error}
	\acro{NPE}{non-parallelity error}
	\acro{HOMO}{highest-occupied \acl{MO}}
	\acro{LUMO}{lowest-unoccupied \acl{MO}}
\end{acronym}


\bibliography{v2rdm_cas_pdft}

\end{document}